\documentclass{iopart}
\usepackage{graphics}
\usepackage{upgreek}
\usepackage{verbatim}
\usepackage{hyperref}

\newcommand{\lessim}{\;\makebox[0pt][l]{\raisebox{-0.1ex}{${}^<$}}%
  \mbox{\raisebox{0.1ex}{${}_\sim$}}\;}
\newcommand{\gtrsim}{\;\makebox[0pt][l]{\raisebox{-0.1ex}{${}^>$}}%
  \mbox{\raisebox{0.1ex}{${}_\sim$}}\;}

\begin{document}

\title[Gravitational-Wave Sensitivity of LILA]{Fundamental Noise and Gravitational-Wave Sensitivity of the Laser Interferometer Lunar Antenna (LILA)}

\author{Teviet Creighton$^1$, Philippe Lognonn\'e$^2$, Mark P. Panning$^3$, James Trippe$^4$, Volker Quetschke$^1$, and Karan Jani$^4$}
\address{$^1$ South Texas Space Science Institute, University of Texas Rio Grande Valley, Brownsville, TX, USA}
\address{$^2$ Université Paris Cité, Institut de Physique du Globe de Paris, CNRS, Paris, France}
\address{$^3$ Jet Propulsion Laboratory, Pasadena, CA, USA}
\address{$^4$ Vanderbilt Lunar Labs Initiative, Vanderbilt University, Nashville, TN, USA}
\ead{teviet.creighton@utrgv.edu}

\date{\today}

\begin{abstract}
  The Earth's Moon presents a uniquely advantageous environment for detecting astrophysical gravitational waves (GWs) in the frequency range of millihertz to decihertz.  Unlike Terrestrial GW detectors,
  the quiet seismic environment of the Moon does not impede detection in this band; in fact the ground motions of the Moon will be excited by GWs, making the Moon a resonant amplifier at low frequencies.  The Laser Interferometer Lunar Antenna (LILA) mission aims to be limited by thermal Brownian noise in its optics across most target frequencies.  
  %For readily achievable design parameters, the initial 5\,km LILA Pioneer detector targets a characteristic strain sensitivity of $\sim10^{-21}$ from 1\,mHz to 100\,mHz, improving by a factor of 10--100 at Lunar normal mode resonances.  A follow-on 40\,km LILA Horizon observatory would further increase GW sensitivity by a factor of 7.
  By taking advantage of the lunar normal mode resonances, we show that the first phase of the mission, LILA Pioneer, achieves the GW sensitivity required to study astrophysical sources through the millihertz to decihertz range. The advanced phase of the mission, LILA Horizon, would increase GW sensitivity to the cosmological horizon in this band.
\end{abstract}

%\maketitle

\section{Introduction}
\label{s:introduction}

In recent years there has been a surge in interest in Lunar science, in parallel with increasing capabilities of private space enterprises and national interest in crewed missions to the Moon.  One of the more intriguing possibilities for Lunar science is in gravitational-wave (GW) astronomy.  Due to its quiet seismic environment and natural hard vacuum, the Moon is potentially a far more productive environment than Earth for operating GW detectors.  Possible detection schemes range from pure seismic accelerometers that treat the Moon as a mechanical resonant GW antenna~\cite{Harms:2021}, to long-baseline interferometers with optical elements fixed on the Moon's surface~\cite{Katsanevas:2020}, to fully-suspended interferometric antennas~\cite{Jani:2020} akin to Terrestrial observatories such as LIGO and Virgo.

The Lunar Interferometer Laser Antenna (LILA)~\cite{Jani:2025} proposes a phased approach to building a laser-based observatory for gravitational-wave detection and Lunar science.  The initial phase \textsf{LILA Pioneer} would have an L-shaped unsuspended laser interferometer in the 3--5\,km range, while a follow-on \textsf{LILA Horizon} would expand to a 40\,km triangular observatory including unsuspended and suspended optical elements.  This paper presents noise and sensitivity estimates for the unsuspended ``strainmeter'' component of these missions.

The LILA Pioneer mission specifically will consist of a primary corner module and secondary end modules, deployed from a single lander and Lunar Terrain Vehicle.  The corner module contains the laser emitter, receiver, and associated electronics and optics (laser frequency stabilization, input and output mode-cleaners).  The end modules are much simpler, containing only a reflector (a steerable mirror or coherent retroreflective metasurface~\cite{Cuesta:2021}) to maximize the return signal.  All modules will have sunshields and insulated enclosures to regulate temperature, and short beamtubes with baffles and electrostatic precipitators to reduce stray light and dust.

The three stations will be operated as a pair of single-arm strainmeters rather than a conventional Michelson interferometer.  The motivation for the L configuration is twofold: (1) It maximizes the effective baseline (total optical path) within the constraints of a maximum sortie range of the Lunar Terrain Vehicle that will be used to deploy the end stations. (2) In the low-frequency regime, where the instrument is predominantly measuring the excitation of Lunar normal modes, the two independent baselines will allow better discrimination of the mode wave pattern and orientation.

As noted, the base strainmeter design will contain no suspension system; all elements will be fixed to a monolithic optical bench anchored to the Lunar surface.  It will not be cryogenic, but will be maintained at a constant temperature similar to the ambient average temperature $\sim300\,\mathrm{K}$.

\section{Noise Sources}
\label{s:noise-sources}

Dominant fundamental noise sources for Terrestrial GW detectors are
typically some combination of seismic noise (ground motion),
suspension thermal noise, thermal noise in optics and coatings, and
quantum readout noise in the laser system.  Other noise sources
include laser phase noise, scattered light, Newtonian (local
gravitational) noise, and technical noise.  We will consider these in
turn.

\subsection{Seismic Noise}

The Earth is a seismically noisy environment, due to geological
processes, along with ground motions produced by waves, wind, and atmospheric pressure loading. Anthropogenic noise can also be observed near cities. 
In particular, there is significant continuous noise at all points on the Earth from
0.05 to 1 Hz caused by the oceans, referred to as microseismic noise \cite{Peterson1993}.
Terrestrial GW detectors isolate their test masses (mirrors) from
these motions through elaborate suspension systems.  By contrast, the
Moon is a seismically quiet environment, with a largely inactive
interior and few surface phenomena that would produce ground movement.
Apollo seismometers did record many transient ``moonquakes'' due to
tidal stress, meteorite impacts, and local thermal changes, but the continuous 
seismic background consistently remained below the noise floor of the instruments, which 
was just above $10^{-10}\mathrm{m}/\mathrm{s}^2/\sqrt{\mathrm{Hz}}$ at its peak sensitivity near 0.5 Hz \cite{Lognonne+2015b, Nunn+2020}, and has been modeled to be significantly lower if driven 
by micrometeorite impacts \cite{Lognonne+2009}.

In the absence of direct measurements of the Lunar seismic background,
we will consider here the sensitivity of an unsuspended strainmeter to gravitational waves without such background.  A later paper~\cite{Lognonne:2025} will explore the use of LILA strainmeters for Lunar geophysics (\textit{selenophysics}).  Furthermore, in the millihertz band, we assume that
\emph{gravitational waves} will themselves excite seismic motions in
excess of any such background; see Section~\ref{s:response-function}.
Thus in each case we treat ground motion as a potential \emph{signal} rather than a noise
source.

\subsection{Thermal Noise}

Lacking suspensions, the proposed strainmeter will similarly lack
suspension thermal noise, so we need only consider Brownian thermal
noise for optical surfaces along the beam path.  A standard
analysis~\cite{Harry:2009} using fluctuation-dissipation theorem gives
the power spectral density $S_x(f)$ in the position $x$ of an optical
surface to be:
\begin{equation}
  \label{eq:psd-thermal-formula}
  S_x(f) = \frac{2k_BT(1-\sigma^2)\phi_\mathrm{eff}}{\pi^{3/2}fwY}\
\end{equation}
where $k_B$ is Boltzmann's constant, $T$ is the temperature, $\sigma$
and $Y$ are the Poisson ratio and Young's modulus respectively of the optical
substrate material, $w$ is the half-width of the beam spot (assumed
Gaussian), and $\phi_\mathrm{eff}$ is the effective loss angle of any
dissipative mechanical processes.  However, this is an idealized case
where dissipation is only through the bulk of the substrate material.
Thermal noise will be dominated by the most mechanically lossy
(dissipative) element contributing to the optical path, which may be
the mirror coatings, mountings, optical bench, etc.  These effects in
practice are usually determined empirically~\cite{Gras:2017} or
through finite-element modeling~\cite{Arai:2018}.
From~\cite{Arai:2018} we treat the following conservative noise
estimate as achievable in practice:
\begin{equation}
  \label{eq:psd-thermal}
  S_x(f) = \left(6\times10^{-32}\mathrm{m}^2/\mathrm{Hz}\right)\left(\frac{T}{300\,\mathrm{K}}\right)\left(\frac{f}{\mathrm{Hz}}\right)^{\!-1}
\end{equation}
Since the basic strainmeter does not employ high-gain optical cavities on its arms, it will be dominated by the narrowest beam size on the lossiest element anywhere on the optical path: this model assumes the beam size is $0.5\,\mathrm{mm}$ (a typical
laser aperture), and the most lossy elements are the
piezoelectric actuators on the mirrors.  For a Lunar mission it would be preferable to use elecro-optic modulators rather than piezoelectrics for beam shaping and phase control, which would be both lower-voltage and mechanically stiffer.  Optimistically, assuming that we can reach thermal noise comparable to the optical coatings in~\cite{Arai:2018}, then our position noise could be as low as:
\begin{equation}
  \label{eq:psd-thermal-optimistic}
  S_x(f) = \left(1.2\times10^{-33}\mathrm{m}^2/\mathrm{Hz}\right)\left(\frac{T}{300\,\mathrm{K}}\right)\left(\frac{f}{\mathrm{Hz}}\right)^{\!-1}
\end{equation}
Because the actual prefactor for a Lunar mission will require numerical or experimental modeling of the final optical system, our noise curves will include it as a free parameter $S_{x,T}(1\,\mathrm{Hz})$ with ``canonical'' values ranging from $1.2\;\mbox{to}\;60\times10^{-33}\mathrm{m}^2/\mathrm{Hz}$.

\subsection{Quantum Noise}

Quantum noise refers to noise resulting from the quantization of the
electromagnetic field of the laser.  At high frequencies this is
called \textit{shot noise}, the limitation of phase readout at the
photodetector due to stochastic reception of individual photons.  This
gives fluctuations in the measured optical path length $x$:
\begin{equation}
  \label{eq:psd-shot}
  S_x(f) = \frac{\hbar c\lambda}{2\pi\eta P}
  = (5.4\times10^{-32}\mathrm{m}^2/\mathrm{Hz})
  \left(\frac{\eta P}{100\,\mathrm{mW}}\right)^{\!-1}
\end{equation}
where $\lambda=1064\,\mathrm{nm}$ is the wavelength of the laser, $P$
is the laser power, and $\eta$ is the quantum efficiency (fraction of
emitted photons detected at the photoreceptor), including losses along
the optical path.  (See e.g.~\cite{LISA:1998} Eq.~(4.4).)

We note that for $\eta P\gtrsim100\,\mathrm{mW}$ (a reasonable readout
power), we will not be dominated by shot noise below 1\,Hz for the more conservative thermal noise estimates.  However, if we pursue the more aggressive thermal noise limits, then we will also want to increase the power on the photoreceptor: $\sim500\,\mathrm{mW}$ would be a practical upper limit.
We also note that rigidly-mounted optics will not be subject to the
low-frequency \textit{radiation pressure noise} that limits the
performance of high-power laser interferometers with suspended optics.

%Notably, shot noise can be reduced by increasing laser power, though
%this increases another quantum noise source: \textit{radiation
%  pressure noise}, due to the fluctuating pressure of individual
%photons hitting the mirrors.  This results in a power spectral density
%scaling as $1/mf^2$ where $m$ is the mass of the optical component.

%In our case, we can neglect radiation pressure noise, since our optics
%are monolithic (effective $m$ is very large).  We only require $\eta
%P$ to be sufficient so that shot noise does not dominate in our target
%band $\lessim1\,\mathrm{Hz}$.  For $\eta P\gtrsim90\,\mathrm{mW}$ we
%can neglect quantum noise, treating thermal noise (above) as the only
%significant noise source below $1\,\mathrm{Hz}$.

\subsection{Other Noise}

We list here a few noise sources that we have considered and believe
can be neglected or mitigated in order to estimate the detector's
ultimate sensitivity.

Stray light from objects other than the distant retroreflector will
introduce noise in range measurements.  However, a well-designed
output mode-cleaner should be able to reject off-axis light, leaving
us susceptible only to dust scattering within the beam path.  The
LADEE mission~\cite{Horanyi:2015} measured dust densities of
$N\sim3\times10^{-3}\mathrm{m}^{-3}$ below 40\,km with particle sizes
$a\lessim0.7\,\upmu\mathrm{m}$.  The relatively uniform density profile below 40\,km suggests typical ballistic grain speeds $v\sim300\,\mathrm{m}/\mathrm{s}$.  Using scattering formulae (12), (13), and~(14)
from~\cite{Cozzumbo:2024} (and adjusting to our notation), the estimated scattering noise is:
\begin{eqnarray}
  \label{eq:scatter}
  S_x(f) &\approx& \frac{16\pi^2(n^2-1)^2}{9}\frac{LNa^6}{wv}
  \nonumber\\
  &\approx&(1.6\times10^{-35}\mathrm{m}^2/\mathrm{Hz})
  \left(\frac{L}{5\,\mathrm{km}}\right)
  \left(\frac{N}{3\times10^{-3}\mathrm{m}^{-3}}\right)\nonumber\\
  &&\;\times\;\left(\frac{a}{0.7\,\upmu\mathrm{m}}\right)^{\!6}
  \left(\frac{v}{300\,\mathrm{m}/\mathrm{s}}\right)^{\!-1}
  \left(\frac{w}{1\,\mathrm{cm}}\right)^{\!-1}
\end{eqnarray}
where $n\approx1.5$ is the grain index of refraction and $w$ is the beam width
along the arm.  This estimate places scattering noise well below shot
noise.

Laser phase noise is often invoked as a reason to use equal-arm interferometers (since it cancels when light from the arms combines), but a frequency-stabilized laser can perform sufficiently for our needs, allowing the two arms to be operated as independent strainmeters. This requires frequency noise of order $d\nu/\nu\sim
dL/L\sim10^{-24}/\sqrt{\mathrm{Hz}}$, or
$d\nu\sim3\times10^{-10}\mathrm{Hz}/\sqrt{\mathrm{Hz}}$, which can be achieved with an on-board high-finesse stabilization cavity.

Newtonian gravitational noise from nearby moving masses is a concern for 3rd Generation Terrestrial detectors at low frequencies.  However, this is because the gravitational coupling bypasses the suspension system that otherwise isolates the test masses from ground motion.  For an unsuspended system the Newtonian gravitational perturbations will always be less than the ground motions, which we are neglecting as a noise source.

``Technical noise'' refers to various measurement errors and perturbations introduced by feedback and control loops, electronic and computational artifacts, etc.  For purposes of this analysis we will treat these as solvable problems rather than immitigable noise.

\section{Response Function}
\label{s:response-function}

Following~\cite{Harms:2022}, we identify two mechanisms by which a gravitational wave may excite a Lunar-based detector: an \textit{inertial regime}, in which the Moon acts like a fluid body, and a \textit{mechanical regime}, in which the Moon behaves like an elastic oscillator.  These two regimes are distinguished by the mechanical response of the Moon $QL_\mathrm{eff}$, where $L_\mathrm{eff}$ is the effective lengthscale that interacts with the gravitational wave (of order the wavelength of a seismic wave of the same frequency), and $Q$ is the Moon's mechanical quality factor at that frequency.  When $QL_\mathrm{eff}\ll R_\mathrm{Moon}$, the Moon behaves inertially; when $QL_\mathrm{eff}\gg R_\mathrm{Moon}$, the Moon behaves resonantly.

\subsection{Lunar Inertial Response}

The inertial regime is perhaps the easiest to understand.  In this limit, the surface responds to the gravitational wave like a collection of free masses, with no proper acceleration in the horizontal direction.  However, the change in distance between a pair of masses separated by a distance $L$ goes as:
\begin{equation}
\delta x \;=\; \mbox{$\frac{1}{2}$}hL
\end{equation}
where $h$ is the projection of the gravitational wave on the arm axis.  For three masses at the corner and ends of two arms with an opening angle $\theta>30^\circ$, we can improve on this by measuring the changing \emph{difference} in arm length, giving:
\begin{equation}
\delta x \;=\; hL\sin\theta
\end{equation}
This is the sensitivity to an \emph{optimally oriented} gravitational wave.  For a single-arm strainmeter, this is a $+$ polarized wave incident at right angles to the axis $L$.  For a two-arm configuration, it is a wave incident from the zenith, whose $+$ polarization axes are symmetric about (i.e.\ have the same bisector as) the two arms.  We note that the ``beam pattern'' response for non-optimally-incident waves is conventionally included in the $h(t)$ of the source object or population rather than the instrument sensitivity curve: the reduction in response is of order $~0.4$, and will differ by tens of percent depending on the beam pattern and assumptions about the source distribution.

The resulting strain sensitivity is simply:
\begin{equation}
S_h(f) \;=\; \frac{S_x(f)}{L^2\sin^2\theta}
\end{equation}
where $\theta\approx90^\circ$ for LILA-Pioneer and $60^\circ$ for LILA-Horizon.  If operating as a single-arm strainmeter we replace $\sin\theta$ with $\frac{1}{2}$.

\subsection{Lunar Mechanical Response}

When $QL_\mathrm{eff} \gg R_\mathrm{Moon}$, the mechanical oscillations excited in the Moon produce proper accelerations whose displacements begin to exceed the relative displacements of free masses.  For frequencies near the Lunar normal modes, this can lead to significantly higher responses.

There has been extensive modeling of this response in the context of seismometer-based detectors, since such detectors \emph{only} respond to proper accelerations.  We take as our starting point the horizontal response function in Figure~5 of~\cite{Bi:2024}.  The precise shape of this response curve is dependent on numerical simulations, based on a plausible but hypothetical Lunar model, so we first consider the broad behaviour rather than the specific detailed structure.

Generally, their base response (which we can identify with $L_\mathrm{eff}$) goes as a power law from $10^3\,\mathrm{m}$ per unit strain at $10^0\,\mathrm{Hz}$, to $10^6\,\mathrm{m}$ per unit strain at $10^{-3}\,\mathrm{Hz}$.  However, the response has peaks at mode frequencies where the Moon acts as a resonant amplifier, increasing the response by a factor $Q$ of up to $10^2$ at $1\,\mathrm{Hz}$, to $10^4$ at $1\,\mathrm{mHz}$.
\begin{equation}
  \label{eq:leff}
  L_\mathrm{eff} \;\approx\; 10^3\,\mathrm{m}\left(\frac{f}{\mathrm{Hz}}\right)^{-1}\;,\qquad
  Q \;\sim\; 10^2\left(\frac{f}{\mathrm{Hz}}\right)^{-2/3}\;.
\end{equation}
Realistically, partial melting in the deep mantle of the Moon will limit the low-frequency $Q$ to $\lessim500$ or so, and scattering due to upper mantle and crust ``megaregolith'' will reduce $Q$ sharply above $\sim0.1\,\mathrm{Hz}$: see Section~\ref{s:gravitational-wave-sensitivity} for the more realistic assumptions used in our final sensitivity curve.

These modes will all have $l=2$ quadrupolar \emph{angular} shape (albeit with different \emph{radial} eigenfunctions depending on frequency).  Since a strainmeter of length $L\ll R_\mathrm{Moon}$ only measures a small fraction of an angular wavelength, the \emph{differential} horizontal displacement will be reduced by a factor $kL$ where $k=(l+\frac{1}{2})R_\mathrm{Moon}$ is the angular wavenumber.  As with the inertial response, a gravitational wave that induces a shear in orthogonal directions will produce a differential displacement up to twice this amount:
\begin{equation}
\label{eq:response-mechanical}
\delta x \;=\; \frac{5QL_\mathrm{eff}L\sin\theta}{R_\mathrm{Moon}} h
\end{equation}
We note that at high frequencies, as $QL_\mathrm{eff}$ decreases, this displacement will eventually be less than the inertial displacement $\delta x=Lh\sin\theta$; this corresponds to the transition to the inertial regime, above.  Thus:
\begin{equation}
S_h(f) \;\approx\; \frac{S_x(f)}{L^2\sin^2\theta}\left[1+\frac{25L_\mathrm{eff}^2 Q^2}{R_\mathrm{Moon}^2}\right]^{-1}
\end{equation}
Of course the response at a normal mode $m$ is not a pure impulse at the resonant frequency $f_m$; even excitation at $f\neq f_m$ may have some resonant enhancement.  To represent this, we replace the $Q$ amplification factor with a bandpass response $T_m(f)$ from each mode:
\begin{eqnarray}
\label{eq:mechanical}
S_h(f) &\approx& \frac{S_x(f)}{L^2\sin^2\theta}\left[1+\frac{25L_\mathrm{eff}^2}{R_\mathrm{Moon}^2}\sum_m T_m(f)\right]^{-1} \\
\label{eq:resonance}
T_m(f) &=& \frac{1}{\left(f_m/f-f/f_m\right)^2+1/Q^2}
\end{eqnarray}

\subsection{Low-Frequency Cutoff}

At very low frequencies, below the Moon's fundamental mode $f_0\approx1\,\mathrm{mHz}$, there should be no inertial regime: we expect the Moon to behave as a solid elastic object rather than a fluid, with displacement proportional to the driving force $\delta x\propto L\nabla g$ where $\nabla g=\frac{1}{2}\ddot{h}=\frac{1}{2}\omega^2 h$ is the tidal gravity gradient.  We model this by removing the fundamental mode response $T_0$ from the sum in Equation~(\ref{eq:response-mechanical}) and instead apply an overall highpass filter $(f/f_0)T_0(f)$ to the response (the inverse of this to the sensitivity), giving the required $\propto f^{-2}$ behaviour for $f\ll f_0$ and flat above $f_0$.
\begin{equation}
\label{eq:response-full}
S_h(f) \approx \frac{S_x(f)}{L^2\sin^2\theta}\frac{f_0/f}{T_0(f)}\left[1+\frac{25L_\mathrm{eff}^2}{R_\mathrm{Moon}^2}\sum_{m\neq0} T_m(f)\right]^{-1}
\end{equation}

\section{Gravitational-Wave Sensitivity}
\label{s:gravitational-wave-sensitivity}

\begin{figure}
\hfill\resizebox{370pt}{!}{\includegraphics{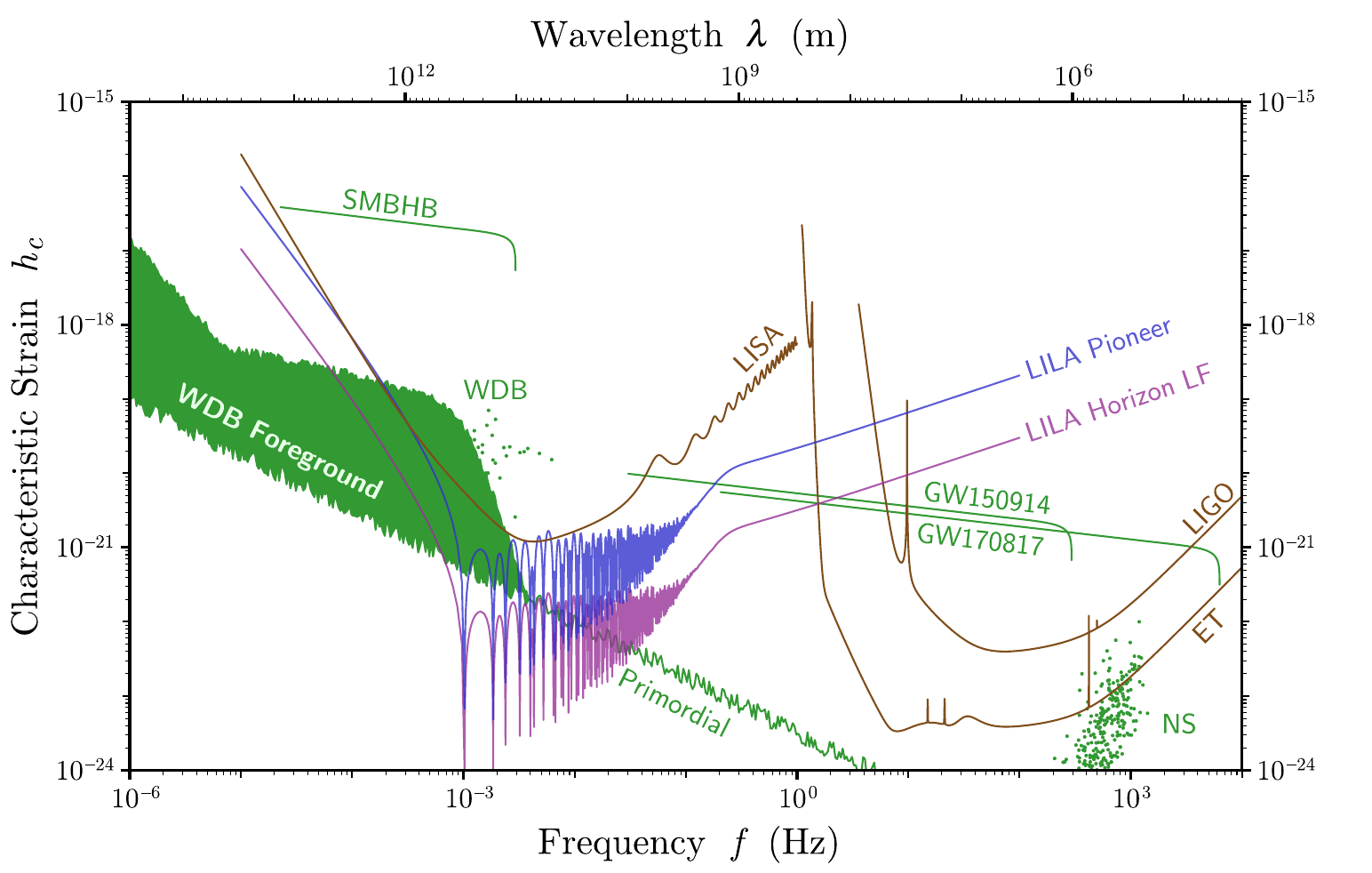}}
\caption{\label{fig:sensitivity}Characteristic strain sensitivity
    for several current and planned gravitational-wave detectors.  The blue and purple curves show the expected sensitivity of the proposed \textsf{LILA Pioneer} (L-shaped 5\,km) and \textsf{LILA Horizon LF} (triangular 40\,km) strainmeters described in the text.}
\end{figure}

Combining Equations~(\ref{eq:psd-thermal-optimistic}), (\ref{eq:psd-shot}), and~(\ref{eq:response-full}), taking $L_\mathrm{eff}=\frac{R_\mathrm{Moon}}{1740}(f/\mathrm{Hz})^{-1}$, we get the following
estimated LILA noise $S_h(f)$ and characteristic sensitivity $h_c=$ \raisebox{0pt}[0pt][0pt]{$\sqrt{fS_h(f)}$}:

\begin{eqnarray}
  S_h &=& \left(4\times10^{-41}/\mathrm{Hz}\right)\left(\frac{L\sin\theta}{5\,\mathrm{km}}\right)^{-2} \nonumber\\
&&\times\left[\left(\frac{S_{x,T}(1\,\mathrm{Hz})}{10^{-33}\mathrm{m}^2/\mathrm{Hz}}\right)\!\left(\frac{T}{300\,\mathrm{K}}\right)\!\left(\frac{f}{\mathrm{Hz}}\right)^{\!-1}\!\!\!+54\left(\frac{\eta
      P}{100\,\mathrm{mW}}\right)^{\!-1}\right] \nonumber\\
  \label{eq:sensitivity-sh}
  &&\times\frac{f_0/f}{T_0(f)}\left[1+(8.3\times10^{-6})\left(\frac{f}{\mathrm{Hz}}\right)^{-2}\sum_{m\neq0}T_m(f)\right]^{-1} \\
%\end{eqnarray}
%\end{samepage}
%
%\noindent Written as a %\emph{characteristic strain} %$h_c=\sqrt{fS_h(f)}$:
%
%\begin{samepage}
%\begin{eqnarray}
  h_c &=& \left(6.3\times10^{-21}\right)\left(\frac{L\sin\theta}{5\,\mathrm{km}}\right)^{-1} \nonumber\\
&&\times\left[\left(\frac{S_{x,T}(1\,\mathrm{Hz})}{10^{-33}\mathrm{m}^2/\mathrm{Hz}}\right)\!\left(\frac{T}{300\,\mathrm{K}}\right)+54\left(\frac{\eta
      P}{100\,\mathrm{mW}}\right)^{\!-1}\!\left(\frac{f}{\mathrm{Hz}}\right)\right]^{1/2} \nonumber\\
  \label{eq:sensitivity-h}
  &&\times\sqrt{\frac{f_0/f}{T_0(f)}}\left[1+(8.3\times10^{-6})\left(\frac{f}{\mathrm{Hz}}\right)^{-2}\sum_{m\neq0}T_m(f)\right]^{-1/2}
  \end{eqnarray}

\noindent Rather than using the estimated $Q\approx100(f/\mathrm{Hz})^{-2/3}$, we take normal mode frequencies $f_m$ and quality factors $Q_m$ based on~\cite{Weber:2011} and calculated with the code~\cite{mineos-v1.0.2}.  Since these results neglect 3d scattering from shallow structures, we impose an exponential cutoff $\sim e^{-f_m/0.05\,\mathrm{Hz}}$ on $Q_m$.
The resulting sensitivity is plotted in Figure~\ref{fig:sensitivity}, under two different design assumptions:
\begin{itemize}
\item\textsf{LILA Pioneer:} The target sensitivity for the initial LILA mission, assuming a 5\,km right-angled ($\theta=90^\circ$) baseline, coating-dominated thermal noise $S_{x,T}(1\,\mathrm{Hz})=1.2\times10^{-33}\mathrm{m}^2/\mathrm{Hz}$, and high-power $500\,\mathrm{mW}$ photodiodes.
\item\textsf{LILA Horizon LF:} The low-frequency component of the LILA Horizon observatory.  As Pioneer but with a 40\,km triangular ($\theta=60^\circ$) baseline.
\end{itemize}
Figure~\ref{fig:sensitivity} also includes for comparison the design sensitivity of the current \textsf{LIGO} detector~\cite{Barsotti:2018}, the planned sensitivities for Einstein Telescope (\textsf{ET})~\cite{Hild:2011} and the Laser Interferometer Space Antenna (\textsf{LISA})~\cite{Babak:2021}, and the characteristic signals from several representative sources:
\begin{itemize}
\item Known neutron stars (\textsf{NS})~\cite{Manchester:2005}, assuming non-axisymmetric ellipticity $10^{-9}$ and one year of integration.
\item Inspiral events \textsf{GW150914} (a black hole binary) and \textsf{GW170817} (a neutron star binary), showing the last year of inspiral.
\item A supermassive black hole binary (\textsf{SMBHB}) of two $3\times10^6M_\odot$ black holes at $10\,\mathrm{Gpc}$, again showing the last year of inspiral.
\item Known white dwarf binaries (\textsf{WDB})~\cite{Kupfer:2024} assuming one year of integration, and an estimated confusion limit (\textsf{WDB Foreground}) from other such binaries in our Galaxy assuming 1/year frequency resolution.
\item A standard Harrison-Zel'dovich (\textsf{Primordial}) gravitational-wave background from the Big Bang singularity.  We note that verification of a stochastic signal will typically require two independent detectors at this sensitivity.
\end{itemize}

\section{Conclusions}
\label{s:conclusions}

This paper has presented basic design parameters and sensitivity estimates for a Lunar gravitational-wave detector using unsuspended laser strainmeters.  The proposed system is based on established technologies already developed for Terrestrial observatories, and can be deployed with modest astronaut participation at arbitrary locations on the Moon.

As shown in Figure~\ref{fig:sensitivity}, by harnessing the power of precision laser ranging along with resonant amplification by the Moon itself, we can achieve sensitivities competitive with orbital gravitational-wave detectors at millihertz frequencies, while also opening an important window into the decihertz band below the range of Terrestrial detectors.  This will facilitate important science goals, from long-duration observation of stellar, intermediate, and supermassive binary systems, to probing the primordial singularity itself.  The Moon promises to be an exciting new platform to revolutionize our understanding of the Universe.

\section*{Acknowledgements}

T.C. and V.Q. were supported by the National Science Foundation grant NSF-2207999. M.P.P. was supported by funds from the Jet Propulsion Laboratory, California Institute of Technology, under a contract with the National Aeronautics and Space Administration (80NM0018D0004). K.J. and J.T. were supported by the Scaling Grant from the Vanderbilt Office of Vice-Provost of Research and Innovation. 

\section*{References}

\bibliography{lilanoise}{}

\end{document}